\documentclass[10pt]{article}
\usepackage{amsmath}
\usepackage{amsfonts}

\usepackage{mathrsfs,amsthm,amsmath,amssymb}
\usepackage{graphicx}
\usepackage{color}
\usepackage{url}

\theoremstyle{definition}
\newtheorem{Theorem}{Theorem}
\newtheorem{Corollary}{Corollary}

\newtheorem{Remark}{Remark}
\newtheorem{Lemma}{Lemma}

\newtheorem{Open}{Problem}

\newcommand{\ftwon}{{\mathbb F}_{2^n}}
\newcommand{\ftwom}{{\mathbb F}_{2^m}}
\newcommand{\ftwo}{{\mathbb F}_{2}}
\newcommand{\Tr}{{\rm {Tr}}}
\newcommand{\Proof}{\noindent\textbf{Proof.}~}

\newcommand{\ls}[1]
    {\dimen0=\fontdimen6\the\font\lineskip=#1\dimen0
     \advance\lineskip.5\fontdimen5\the\font
     \advance\lineskip-\dimen0
     \lineskiplimit=0.9\lineskip
     \baselineskip=\lineskip
     \advance\baselineskip\dimen0
     \normallineskip\lineskip\normallineskiplimit\lineskiplimit
     \normalbaselineskip\baselineskip
     \ignorespaces}

\begin{document}

\bibliographystyle{abbrv}

\title{Several Classes of Permutation Trinomials From Niho Exponents}
\author{ Nian Li and Tor Helleseth
\thanks{The authors are with the Department of Informatics, University of Bergen,
 N-5020 Bergen, Norway. Email: Nian.Li@ii.uib.no;
Tor.Helleseth@ii.uib.no.}
}
\date{}
\maketitle
\ls{1.5}

\thispagestyle{plain} \setcounter{page}{1}

\begin{abstract}
Motivated by recent results on the constructions of permutation polynomials with few terms over the finite field $\ftwon$, where $n$ is a positive even integer, we focus on the construction of permutation trinomials over $\ftwon$ from Niho exponents. As a consequence, several new classes of permutation trinomials over $\ftwon$ are constructed from Niho exponents based on some subtle manipulation of solving equations with low degrees over finite fields.
\end{abstract}

\section{Introduction}

A permutation polynomial over a finite field is a polynomial that acts as a permutation of the elements of the field. Permutation polynomials were first studied by Hermite \cite{Hermite} for the case of finite prime fields and by Dickson \cite{Dickson} for arbitrary finite fields. Permutation polynomials have wide applications in many areas of mathematics and engineering such as coding theory, cryptography and combinatorial designs. Permutation polynomials over finite field $\ftwon$ without further considerations are not difficult to construct since the number of permutation polynomials over $\ftwon$ is $2^n!$ and all of which can be obtained from the Lagrange interpolation. However, researchers are only interested in the permutation polynomials that either have a simple algebraic form or possess additional extraordinary properties which are required by their applications in mathematics and engineering \cite{Hou}. In this sense it is normally difficult to construct permutation polynomials. The reader is referred to the survey paper \cite{Hou} for recent developments on the constructions of permutation polynomials.

The construction of permutation polynomials with few terms attracts researchers' interest in recent years due to their simple form \cite{Ding-QWYY,Hou14,Hou15,Li-Qu-Chen,Tu-Zeng-Hu-Li,Tu-Zeng-Hu,ZTT,ZZC,Zieve-subgroup} and some additional nice properties, for example, their close relations with the Helleseth's $-1$ conjecture \cite{Helleseth}, Welch's conjecture \cite{Dobbertin} and Niho's conjecture \cite{Dobbertin-Nihocase}. Currently, only a small number of specific classes of permutation binomials and trinomials are described in the literature, see the survey paper \cite{Hou13} and two recent papers \cite{Ding-QWYY,Li-Qu-Chen} on permutation binomials and trinomials over finite fields. Tu et al. in \cite{Tu-Zeng-Hu-Li} investigated the permutation property of a class of polynomials over $\ftwon$ of the form
 \begin{eqnarray}\label{def-Tu}
   \sum_{i=1}^k a_ix^{s_i(2^m-1)+e}, a_i\in\ftwon
 \end{eqnarray}
by using certain techniques in calculating exponential sums, where $n=2m$, $k, s_i, e$ are integers. They proved that the polynomial is a permutation over $\ftwon$ if and only if a related equation has a unique solution in the unit circle of $\ftwon$. Tu et al. obtained some classes of permutation binomials and monomial complete permutation polynomials in \cite{Tu-Zeng-Hu-Li} for $k=2$. However, for the case of $k\ge 3$, as pointed out by Tu et al. themselves, it is difficult to derive new permutation polynomials from their method since it is a challenging problem to determine the solutions to the corresponding equation if $k\ge 3$. Based on this fact and their computer experiments, Tu et al. proposed a conjecture on two explicit permutation trinomials of the form \eqref{def-Tu} in \cite{Tu-Zeng-Hu-Li}. Motivated by the results in \cite{Tu-Zeng-Hu-Li}, Zieve considered permutation polynomials from a different approach in \cite{Zieve-subgroup}. By exhibiting classes of low-degree rational functions over finite fields, Zieve solved the conjecture proposed in \cite{Tu-Zeng-Hu-Li} and also constructed some permutation polynomials.

Inspired by their work, in this paper we concentrate on the construction of permutation trinomials over $\ftwon$ from Niho exponents of the form
\begin{eqnarray}\label{def-PP}
  f(x)=x+x^{s(2^m-1)+1}+x^{t(2^m-1)+1},
\end{eqnarray}
where $n=2m$ and $1\leq s, t\leq 2^m$. A positive integer $d$ is called a {\em Niho exponent} with respect to the finite field $\ftwon$ if $d \equiv 2^j\pmod{2^m-1}$ for some nonnegative integer $j$. When $j=0$, the integer $d$ is then called a normalized Niho exponent. The Niho exponents were originally introduced by Niho \cite{Niho-PhD} who investigated the cross-correlation between an $m$-sequence and its $d$-decimation. We aim to find new pairs for $(s,t)$ such that $f(x)$ defined by \eqref{def-PP} is a permutation polynomial. As a consequence, several new classes of permutation trinomials over $\ftwon$ with the form \eqref{def-PP} are constructed based on some subtle manipulation of solving equations with low degrees over finite fields.

\section{Preliminaries} %

Throughout this paper, let $n=2m$ and $\ftwon$ denote the finite field with $2^n$ elements. Denote the conjugate of $x\in\ftwon$ over $\ftwom$ by $\overline{x}$, i.e., $\overline{x}=x^{2^m}$. The unit circle of $\ftwon$ is defined as follows:
\[U=\{x\in\ftwon: x\overline{x}=x^{2^m+1}=1\}.\]

The unit circle of $\ftwon$ has the following relation with the finite field $\ftwom$.

\begin{Lemma}(\cite{Lahtonen}) \label{lem-U}
  Let $\gamma$ be any fixed element in $\ftwon\backslash\ftwom$. Then we have
   \begin{eqnarray*}
   U\backslash\{1\}=\{\frac{z+\gamma}{z+\overline{\gamma}}: z \in \ftwom\}.
       \end{eqnarray*}
\end{Lemma}

To prove our main results, in the sequel we will consider the number of solutions to certain equations with low degrees over finite fields. It will be seen that the following two lemmas play significant roles in the proofs of our main results.

\begin{Lemma}(\cite{Lidl-N}) \label{lem-Quadratic}
   Let $a,b\in\ftwon$ with $a\ne 0$. Then $x^2+ax+b=0$ has solution in $\ftwon$ if and only if $\Tr_1^n(b/a^2)=0$, where $\Tr_1^n(\cdot)$ is the trace function from $\ftwon$ to $\ftwo$.
\end{Lemma}

Leonard and Williams \cite{Leonard-W} described the factorization of a quartic polynomial over the finite field $\ftwom$ in terms of the roots of a related cubic polynomial. In the sequel we only need the result which was characterized in the points $(b)$ and $(e)$ of their main theorem.

\begin{Lemma}\label{lem-Leonard-W}\cite{Leonard-W}
 Let $h(x)=x^4+a_2x^2+a_1x+a_0\in\ftwom[x]$ with $a_0a_1\ne 0$ and $g(y)=y^3+a_2y+a_1$. Let $r_i$ be the roots of $g(y)=0$ when they exist in $\ftwom$ and $w_i=a_0r_i^2/a_1^2$ for $i=1,2,3$. Then $h(x)$ has no solution in $\ftwom$ if one of the following conditions is satisfied:
 \begin{enumerate}
   \item [1)] $g(y)=0$ has exactly one solution in $\ftwom$ and $\Tr_1^m(w_1)=1$; or
   \item [2)] $g(y)=0$ has exactly three solutions in $\ftwom$ and $\Tr_1^m(w_1)=0$, $\Tr_1^m(w_2)=\Tr_1^m(w_3)=1$.
\end{enumerate}
\end{Lemma}

Notice that the inverse of a normalized Niho exponent, if exists, is still a normalized Niho exponent and the product of two normalized Niho exponents is also a normalized Niho exponent. This implies the following lemma.
For simplicity, if the integers $s, t$ are written as fractions, then they should be interpreted as modulo $2^m+1$. For instance, $(s,t)=(\frac{1}{2},\frac{3}{4})=(2^{m-1}+1,2^{m-2}+1)$.

\begin{Lemma}\label{lem-equiv}
Let $(s,t)=(i,j)$ be a pair such that $f(x)$ defined by \eqref{def-PP} is a permutation polynomial, then $f(x)$ defined by \eqref{def-PP} is also a permutation polynomial for the following pairs
\begin{enumerate}
  \item [1)] $(s,t)=(\frac{i}{2i-1},\frac{i-j}{2i-1})$ if $\gcd(2i-1,2^m+1)=1$;
  \item [2)] $(s,t)=(\frac{j}{2j-1},\frac{j-i}{2j-1})$ if $\gcd(2j-1,2^m+1)=1$.
\end{enumerate}
\end{Lemma}

\Proof  We only prove the case for $\gcd(2i-1,2^m+1)=1$ since the other case can be proved in the same manner. Let $d_i=i(2^m-1)+1$ and $d_j=j(2^m-1)+1$, by $\gcd(2i-1,2^m+1)=1$, we have $\gcd(d_i,2^n-1)=1$, i.e., the inverse of $d_i$ exists. This implies that $f(x^{d_i^{-1}})$ is also a permutation if $f(x)$ is a permutation. Then, the results follows from the fact $d_i^{-1}=\frac{i}{2i-1}(2^m-1)+1$ and $d_i^{-1}d_j=\frac{i-j}{2i-1}(2^m-1)+1$. This completes the proof.   \hfill $\Box$

The following lemma, which will be used to prove our main results, was proved by Park and Lee in 2001 and reproved by Zieve in 2009.

 \begin{Lemma}(\cite{Park-Lee,Zieve-09}) \label{lem-Zieve}
  Let $p$ be a prime, $n$ a positive integer and $h(x)\in \mathbb{F}_{p^n}[x]$. If $d, s, r>0$ such that $p^n-1=ds$, then $x^rh(x^s)$ is a permutation over $\mathbb{F}_{p^n}$ if and only if
 \begin{enumerate}
   \item [1)] $\gcd(r,s)=1$, and
   \item [2)] $x^rh(x)^s$ permutes the $d$-th roots of unity in $\mathbb{F}_{p^n}$.
 \end{enumerate}
\end{Lemma}

The polynomials of the form $x^rh(x)^s$ have been widely studied, see \cite{Akbary-Wang,Park-Lee,Tu-Zeng-Hu-Li,Wan-Lidl,Zieve-subgroup,Zieve-09} for example. The permutation property of this type of polynomials has been characterized as Lemma \ref{lem-Zieve} in a simple form. Lemma \ref{lem-Zieve} reduces the problem of determination of permutations over $\mathbb{F}_{p^n}$ to that of determination of permutations over the subgroups of $\mathbb{F}_{p^n}$. However, it is still a difficult problem to verify the second condition in Lemma \ref{lem-Zieve}.

\section{Known Permutation Trinomials over $\ftwon$}

Except the linearized permutation trinomials described in \cite{Lidl-N} and the Dickson polynomial of degree seven, the following is a list of known classes of permutations trinomials over $\mathbb{F}_{2^{n}}=\mathbb{F}_{2^{2m}}$:

\begin{Theorem} \label{thm-known}
The known classes of permutation trinomials over $\mathbb{F}_{2^{n}}=\mathbb{F}_{2^{2m}}$ are:
\begin{enumerate}
    \item [1)] $f_1(x)=x^{2^{2k}+1}+(ax)^{2^k+1}+ax^2$, where $2m=3k$ and $a^{2^{2k}+2^k+1}\ne 1$ \cite{Blokhuis-CHK}.
    \item [2)] $f_2(x)=x^{2^{2k}+1}+x^{2^k+1}+vx$, where $2m=3k$ and $0\ne v\in\mathbb{F}_{2^k}$ \cite{Tu-Zeng-Hu}.
    \item [3)] $f_3(x)=x^rh(x^s)$, where $s=\frac{2^{2m}-1}{3}$, $h(x)=ax^2+bx+c$, $\alpha$ is a generator of $\mathbb{F}_{2^{2m}}$, $w=\alpha^{s}$, $\gcd(r,s)=1$, $h(w^i)\ne 0$ for $i=0,1,2$, and $\log_{\alpha}(h(1)/h(w))\equiv \log_{\alpha}(h(w)/h(w^2))\not\equiv r\pmod{3}$  \cite{Lee-Park}.
    \item [4)] $f_4(x)=x^{k(2^m+1)+3}+x^{k(2^m+1)+2^m+2}+x^{k(2^m+1)+3\cdot 2^m}$, where $k$ is a nonnegative integer and $\gcd(2k+3,2^m-1)=1$  \cite{Zieve-subgroup}.
    \item [5)] $f_5(x)=x+ax^{2(2^m-1)+1}+a^{2^{m-1}}x^{2^m(2^m-1)+1}$, where $a^{2^m+1}=1$  \cite{Ding-QWYY, Li-Qu-Chen}.
    \item [6)] $f_6(x)=x+x^{k(2^m-1)+1}+x^{-k(2^m-1)+1}$, where $k$ is a positive integer, either $m$ is even or $m$ is odd and $\exp_3(k)\ge \exp_3(2^m+1)$. Here $\exp_3(k)$ denotes the exponent of 3 in the canonical factorization of $k$ \cite{Ding-QWYY}.
     \item [7)] $f_7(x)=ax+bx^{2^m}+x^{2(2^m-1)+1}$, where $ab\ne0$, either $a=b^{1-2^m}$, $\Tr_1^m(b^{-1-2^m})=0$ or $a\ne b^{1-2^m}$, $ab^{-2}\in\mathbb{F}_{2^{m}}$, $\Tr_1^m(ab^{-2})=0$ and $b^2+a^2b^{2^m-1}+a=0$ \cite{Hou14,Hou15}.
    \item [8)] $f_8(x)=x+x^{2^m}+x^{2^{m-1}(2^m-1)+1}$, where $m\not\equiv 0 \pmod{3}$ \cite{Li-Qu-Chen}.
    \item [9)] $f_9(x)=x+x^{2^m+2}+x^{2^{m-1}(2^m+1)+1}$, where $m$ is odd  \cite{Li-Qu-Chen}.
  \end{enumerate}
\end{Theorem}

\begin{Remark} For the above known permutation trinomials we find that:
\begin{enumerate}
  \item [1)] The polynomials $f_3(x),f_4(x),f_6(x)$ can also be permutations over certain finite fields with odd characteristic $p$, see \cite{Lee-Park,Zieve-subgroup,Ding-QWYY} respectively.
  \item [2)] The permutation $f_5(x)$ is due to Ding et al. for $a=1$ \cite{Ding-QWYY} and Li et al. for $a^{2^m+1}=1$ \cite{Li-Qu-Chen}. Notice that $f_5(x)$ is a permutation if and only if $x(1+ax^2+a^{2^{m-1}}x^{2^m})^{2^m-1}$ permutes the unit circle $U$ of $\ftwon$ according to Lemma \ref{lem-Zieve}. Let $y=a^{2^{n-1}}x$, then $ax^2=y^2$ and $a^{2^{m-1}}x^{2^m}=(ax^2)^{2^{m-1}}=y^{2^m}$. This implies that $x(1+ax^2+a^{2^{m-1}}x^{2^m})^{2^m-1}$ permutes $U$ if and only if $a^{-2^{n-1}}y(1+y^2+y^{2^m})^{2^m-1}$ permutes $U$. Thus, the generalization of $f_5(x)$ in \cite{Li-Qu-Chen} from $a=1$ to $a^{2^m+1}=1$ is trivial.
  \item [3)] Tu et al. first conjectured in \cite{Tu-Zeng-Hu-Li} that $g(x):=x^{2^m}+x^{2(2^m-1)+1}+x^{2^m(2^m-1)+1}$ is a permutation over $\mathbb{F}_{2^{2m}}$ if $m$ is odd. Zieve solved the conjecture based on low-degree rational functions over finite fields
      \cite[Corollary 1.4]{Zieve-subgroup} and showed that $g(x)$ can be obtained from $f_4(x^{2^n-2})$ with $k=2^m-3$. Notice that $g(x)^{2^m}=f_5(x)$ with $a=1$. Then $f_5(x)$ with $a=1$ is covered by $f_4(x)$.
\end{enumerate}
\end{Remark}

In fact, the polynomials $f_4(x)$ in Theorem \ref{thm-known} and $f(x)$ defined by \eqref{def-PP} have the following relation. It should be noticed that all the three exponents of $f_4(x)$ in Theorem \ref{thm-known} equal to $(2k+3)$ modulo $(2^m-1)$. Thus, for any $d$ with $\gcd(d,2^n-1)=1$ and $0\le i\le n-1$, if $(2k+3)d\equiv 2^i\pmod{2^m-1}$, then it can be seen that $f_4(x^d)^{2^{n-i}}=\sum_{j=1}^3x^{d_j}$, where $d_j=s_j(2^m-1)+1$ for $j=1,2,3$. Moreover, by $(2k+3)d\equiv 2^i\pmod{2^m-1}$ we have $dk=2^{m-1}(2^i-3d)$ and then $(2dk+3d-2^i)/(2^m-1)=(2^i-3d)$ which leads to $s_1=2^{n-i}(dk+(2dk+3d-2^i)/(2^m-1))=2^{n-i}(2^{m-1}+1)(2^i-3d)=2^{n-i-1}(2^i-3d)\pmod{2^m+1}$. Similarly, we can obtain $s_2=2^{n-i-1}(2^i-d)\pmod{2^m+1}$ and $s_3=2^{n-i-1}(2^i+3d)\pmod{2^m+1}$. Observe that $\gcd(d_2,2^n-1)=\gcd(2s_2-1,2^m+1)=\gcd(-2^{n-i}d,2^m+1)=1$, i.e., $d_2^{-1}$ exists. Since the inverse of a normalized Niho exponent is still a normalized Niho exponent and the product of two normalized Niho exponents is also a normalized Niho exponent, let $d_2^{-1}=s_2^{-1}(2^m-1)+1$, $d_1d_2^{-1}=s_{12}(2^m-1)+1$ and $d_3d_2^{-1}=s_{32}(2^m-1)+1$, then we can get $s_2^{-1}=(d-2^i)/2d \pmod{2^m+1}$ and $s_{12}=(s_1+s_2^{-1}-2s_1s_2^{-1})\pmod{2^m+1}=2^{n-i-1}(2^i-3d)(1-(d-2^i)/d)+(d-2^i)/2d=2^{n-1}(2^i-3d)/d+(d-2^i)/2d=(2^i-3d)/2d+(d-2^i)/2d=-1=2^m\pmod{2^m+1}$.
Similarly, we have $s_{32}=2\pmod{2^m+1}$. Therefore, we obtain that
\begin{eqnarray*}
f_4(x^{dd_2^{-1}})^{2^{n-i}}=\sum_{j=1}^3x^{d_jd_2^{-1}}=x+x^{2(2^m-1)+1}+x^{2^m(2^m-1)+1}.
\end{eqnarray*}

Based on Theorem \ref{thm-known} and the above discussion, we immediately have

\begin{Theorem}\label{thm-known pairs}
The known pairs $(s,t)$ such that the trinomials of the form \eqref{def-PP} are permutations over $\mathbb{F}_{2^{2m}}$ are:
\begin{enumerate}
  \item [1)] $(s,t)=(k,-k)$, where $k$ is a positive integer, either $m$ is even or $m$ is odd and $\exp_3(k)\ge \exp_3(2^m+1)$.
  \item [2)] $(s,t)=(2,2^m)=(2,-1)$ for every positive integer $m$.
  \item [3)] $(s,t)=(1,2^{m-1})=(1,-\frac{1}{2})$, where $m\not\equiv 0 \pmod{3}$.
\end{enumerate}
\end{Theorem}

Note that the polynomial $f_7(x)$ in Theorem \ref{thm-known} with $a=b=1$ is also of the form \eqref{def-PP} with $(s,t)=(1,2)$. But this special case is covered by the case of $(s,t)=(k,-k)$, as shown in the following remark.

\begin{Remark} By Lemma \ref{lem-equiv} and Theorem \ref{thm-known pairs}, we can conclude that the following pairs also lead to permutation polynomials of the form \eqref{def-PP} respectively:
\begin{enumerate}
  \item [1)] $(s,t)=(\frac{k}{2k-1},\frac{2k}{2k-1})$ if $\gcd(2k-1,2^m+1)=1$ and the conditions in Theorem \ref{thm-known pairs}-1) are met; and $(s,t)=(\frac{k}{2k+1},\frac{2k}{2k+1})$ if $\gcd(2k+1,2^m+1)=1$ and the conditions in Theorem \ref{thm-known pairs}-1) are met.
  \item [2)] $(s,t)=(1,\frac{1}{3})$ and $(s,t)=(1,\frac{2}{3})$ if $\gcd(3,2^m+1)=1$.
  \item [3)] $(s,t)=(1,\frac{3}{2})$ and $(s,t)=(\frac{1}{4},\frac{3}{4})$ if $m\not\equiv 0 \pmod{3}$.
\end{enumerate}
\end{Remark}

\section{New Permutation Trinomials of the Form \eqref{def-PP}}

 In this section, we present several new classes of permutation trinomials of the form \eqref{def-PP} by using Lemma \ref{lem-Zieve} and some techniques in solving equations with low degrees over the finite field $\ftwon$.

\begin{Theorem}\label{thm-3-14}
 Let $n=2m$ for an even integer $m$. Then the trinomial $f(x)$ defined by \eqref{def-PP} is a permutation if $(s,t)=(-\frac{1}{3},\frac{4}{3})=(\frac{2^{m+1}+1}{3},\frac{2^m+5}{3})$.
\end{Theorem}

\Proof To complete this proof,  according to Lemma \ref{lem-Zieve} we need to show $\phi_1(x)=x(1+x^s+x^t)^{2^m-1}$ permutes the unite circle $U$ of $\ftwon$ which is equivalent to showing $\phi_1(x^3)=x^3(1+x^{3s}+x^{3t})^{2^m-1}=x^3(1+x^{-1}+x^{4})^{2^m-1}$ permutes $U$ since $\gcd(3,2^m+1)=1$. Notice that $x^5+x+1\ne 0$ for any $x\in U$. Otherwise, we have $x^5+x+1=x^5+x+x^{2^m+1}=0$, i.e., $x^{2^m}=x^4+1$ which leads to $x=(x^4+1)^{2^m}=(x^{2^m}+1)^4=x^{16}$. That is, $x^{15}=(x^5)^3=(x+1)^3=1$ which is impossible since $x^2+x+1=0$ has no solution in $U$ when $m$ is even. Thus, $\phi_1(x^3)$ for $x\in U$ can be expressed as
\begin{eqnarray*}
  \phi_1(x^3)=x^3(1+x^{-1}+x^4)^{2^m-1}=\frac{x^5+x^4+1}{x^5+x+1}.
\end{eqnarray*}
Observe that $\gcd(x^5+x^4+1,x^5+x+1)=\gcd(x^4+x,x^5+x+1)=\gcd(x^3+1,x^5+x+1)=\gcd(x^3+1,x^2+x+1)=x^2+x+1$. This means that $\phi_1(x^3)$ can be reduced to
\begin{eqnarray*}
  \phi_1(x^3)=\frac{x^5+x^4+1}{x^5+x+1}=\frac{x^3+x+1}{x^3+x^2+1}.
\end{eqnarray*}
Assume that $x,y\in U$ and $\phi_1(x^3)=\phi_1(y^3)$, in the following we show that $\phi_1(x^3)=\phi_1(y^3)$ if and only if $x=y$. A direct calculation from $\phi_1(x^3)=\phi_1(y^3)$ gives
 \begin{eqnarray*}
 (x+y)[(xy+1)(x+y)+(x^2y^2+xy+1)]=0.
 \end{eqnarray*}
Then we need to show $(xy+1)(x+y)+(x^2y^2+xy+1)=0$ has no solution for $x,y\in U$ with $x\ne y$. Let $y=ux$ for some $1\ne u\in U$, we have
 \begin{eqnarray} \label{eq-u-rd}
(x^4+x^3)u^2+(x^3+x^2+x)u+(x+1)=0
 \end{eqnarray}
and we need to prove that \eqref{eq-u-rd} has no solution $1\ne u\in U$. Notice that $x\ne 1$ otherwise we have $u=0$ and then $y=0$, a contradiction. Then \eqref{eq-u-rd} can be written as
 \begin{eqnarray}\label{eq-u-square}
 u^2+\alpha u+\beta=0,
 \end{eqnarray}
where $\alpha=\frac{x^2+x+1}{x^2(x+1)}$ and $\beta=\frac{1}{x^3}$. Observe that $\frac{\beta}{\alpha^2}=\frac{x(x+1)^2}{(x^2+x+1)^2}\in\ftwom$ due to $x\in U$ and
$$\Tr_1^m(\frac{\beta}{\alpha^2})=\Tr_1^m(\frac{x(x^2+x+1)}{(x^2+x+1)^2}+\frac{x^2}{(x^2+x+1)^2})=0.$$
This together with \eqref{eq-u-square} implies that
\begin{eqnarray*}
 0=\sum_{i=0}^{m-1}((\frac{u}{\alpha})^2+\frac{u}{\alpha}+\frac{\beta}{\alpha^2})^{2^i}=(\frac{u}{\alpha})^{2^m}+\frac{u}{\alpha}
\end{eqnarray*}
which leads to
\begin{eqnarray*}
 u^2=\alpha^{1-2^m}=\frac{1}{x^3}=\beta
\end{eqnarray*}
since $u\in U$. Again by \eqref{eq-u-square} we get $\alpha u=0$, i.e., $x^2+x+1=0$. This is impossible since $x^2+x+1=0$ has no solution in $U$ if $m$ is even. Thus \eqref{eq-u-rd} has no solution $u\in U$. This completes the proof. \hfill $\Box$

\begin{Theorem}\label{thm-3q}
 Let $n=2m$ for an even integer $m$. Then the trinomial $f(x)$ defined by \eqref{def-PP} is a permutation if $(s,t)=(3,-1)=(3,2^m)$.
\end{Theorem}

\Proof To complete the proof, according to Lemma \ref{lem-Zieve}, it is sufficient to show that $\phi_2(x)=x(1+x^3+x^{2^m})^{2^m-1}$ permutes the unit circle $U$ of $\ftwon$. We first show that $x^4+x+1\ne 0$ for any $x\in U$. Note that $x^4+x+1=0$ and $x^{2^m+1}=1$ lead to $x^{2^m}=x^3+1$. Then, $x=(x^3+1)^{2^m}=((x^{2^m})^3+1)=((x^3+1)^3+1)=x^9+x^6+x^3=x\cdot(x^4)^2+x^2\cdot x^4+x^3$ which together with $x^4+x+1=0$ implies that $x=0, 1$, a contradiction with $x^4+x+1=0$ and $x^{2^m+1}=1$. Therefore, $x^4+x+1\ne 0$ and we can rewrite $\phi_2(x)$ as $\phi_2(x)=\frac{x^4+x^3+1}{x(x^4+x+1)}$ for any $x\in U$.

 Assume that $x,y\in U$ and $\phi_2(x)=\phi_2(y)$, then we need to show that it holds if and only if $x=y$. A direct calculation from $\phi_2(x)=\phi_2(y)$ gives
 \begin{eqnarray*}
(x+y)[x^4y^4+(x^3y^3+x^2y^2+xy)(x+y)+x^2y^2+x^4+y^4+x+y+1]=0.
 \end{eqnarray*}
Thus we need to prove that $x^4y^4+(x^3y^3+x^2y^2+xy)(x+y)+x^2y^2+x^4+y^4+x+y+1=0$ has no solution for $x, y\in U$ with $x\ne y$. Let $y=ux$ for some $1\ne u\in U$, then it can be expressed as
 \begin{eqnarray} \label{eq-u-2nd}
 c_4u^4+c_3u^3+c_2u^2+c_1u+c_0=0,
 \end{eqnarray}
where the coefficients are given by
\begin{eqnarray*}
  c_4=(x^8+x^7+x^4),  c_3=(x^7+x^5),  c_2=(x^5+x^4+x^3),  c_1=(x^3+x), c_0=(x^4+x+1).
\end{eqnarray*}
We next show that \eqref{eq-u-2nd} has no solution  $u\in U\backslash\{1\}$ for any $x\in U$.
Applying Lemma \ref{lem-U} to \eqref{eq-u-2nd}, then our problem becomes to show
\begin{eqnarray*}
 c_4(\frac{z+\gamma}{z+\overline{\gamma}})^4+c_3(\frac{z+\gamma}{z+\overline{\gamma}})^3+c_2(\frac{z+\gamma}{z+\overline{\gamma}})^2+c_1(\frac{z+\gamma}{z+\overline{\gamma}})+c_0=0
\end{eqnarray*}
has no solution $z\in\ftwom$ for a fixed $\gamma\in\ftwon\backslash\ftwom$. If $x=1$, the $\eqref{eq-u-2nd}$ is reduced to $u^4+u^2+1=0$ which has no solution in $U$ due to $m$ is even. For any $x\in U$ differs from $1$, taking $\gamma=1/x\in\ftwon\backslash\ftwom$ and multiplying $(z+\overline{\gamma})^4$ on both sides, then the above equation becomes
\begin{eqnarray*}
z^4+\frac{x^8+1}{x^8+x^4+1}z^3+\frac{x^6+x^2}{x^8+x^4+1}z^2+1=0.
\end{eqnarray*}
Replacing $z$ by $1/z$ and multiplying by $z^4$, then, to complete the proof, it suffices to prove that
\begin{eqnarray}\label{eq-z-noroot}
z^4+\frac{x^6+x^2}{x^8+x^4+1}z^2+\frac{x^8+1}{x^8+x^4+1}z+1=0
\end{eqnarray}
has no solution in $z\in\ftwom$ for any $x\in U$ with $x\ne 1$. Then the result follows from Lemma \ref{lem-2nd-4th} presented below.
This completes the proof. \hfill $\Box$

\begin{Lemma}\label{lem-2nd-4th}
 Let the notation be defined as above. Then \eqref{eq-z-noroot} has no solution in $z\in\ftwom$ for any $1\ne x\in U$.
\end{Lemma}

\Proof According to Lemma \ref{lem-Leonard-W}, we need to consider the number of solutions to $g(y)=y^3+a_2y+a_1=0$, where $a_2=\frac{x^6+x^2}{x^8+x^4+1}$, $a_1=\frac{x^8+1}{x^8+x^4+1}$.
For any $1\ne x\in U$, let $c=x+\frac{1}{x}\in\ftwom$, then $x, \frac{1}{x}\in\ftwon\backslash\ftwom$ are the two solutions of $z^2+cz+1=0$ which means that $\Tr_1^m(\frac{1}{c})=1$ due to Lemma \ref{lem-Quadratic}. Moreover, one can obtain
$$a_2=\frac{x^6+x^2}{x^8+x^4+1}=\frac{x^2+x^{-2}}{x^4+x^{-4}+1}=\frac{c^2}{c^4+1},\;a_1=\frac{x^8+1}{x^8+x^4+1}=\frac{x^4+x^{-4}}{x^4+x^{-4}+1}=\frac{c^4}{c^4+1}.$$
Then, $g(y)=0$ can be equivalently written as
\begin{eqnarray*}
(\frac{y}{\frac{c}{c^2+1}})^3+\frac{y}{\frac{c}{c^2+1}}+c^3+c=0,
\end{eqnarray*}
which implies that $\frac{y}{\frac{c}{c^2+1}}=c$, i.e., $y=\frac{c^2}{c^2+1}\in\ftwom$ is a solution to $g(y)=0$. Then, it can be readily verified that the polynomial $g(y)$ can be factorized as
\begin{eqnarray*}
g(y)=(y+\frac{c^2}{c^2+1})(y^2+\frac{c^2}{c^2+1}y+\frac{c^2}{c^2+1}).
\end{eqnarray*}
Note that $\Tr_1^m(\frac{c^2+1}{c^2})=\Tr_1^m(1)+\Tr_1^m(\frac{1}{c})=1$ due to $m$ is even. This implies that $y^2+\frac{c^2}{c^2+1}y+\frac{c^2}{c^2+1}$ has no solution in $\ftwom$ according to Lemma \ref{lem-Quadratic}. Therefore, $g(y)=0$ has exactly one solution $y=r_1=\frac{c^2}{c^2+1}\in\ftwom$, and then the desired result follows from Lemma \ref{lem-Leonard-W} and the fact that $\Tr_1^m(w_1)=\Tr_1^m(r_1^2/a_1^2)=\Tr_1^m(\frac{c^4+1}{c^4})=\Tr_1^m(1)+\Tr_1^m(\frac{1}{c})=1$. This completes the proof. \hfill $\Box$

\begin{Theorem}\label{thm-3-25}
 Let $n=2m$ for an even integer $m$. Then the trinomial $f(x)$ defined by \eqref{def-PP} is a permutation if $(s,t)=(-\frac{2}{3},\frac{5}{3})=(\frac{2^{m}-1}{3},\frac{2^{m+1}+7}{3})$.
\end{Theorem}

\Proof Note that $\gcd(3,2^m+1)=1$ due to $m$ is even. Then, similar as the proof of Theorem \ref{thm-3-14}, to complete this proof, it is sufficient to prove that $\phi_3(x^3)=x^3(1+x^{3s}+x^{3t})^{2^m-1}=x^3(1+x^{-2}+x^5)^{2^m-1}$ permutes $U$. We first need to show $x^7+x^2+1\ne 0$ for any $x\in U$. If $x^7+x^2+1=0$ for some $x\in U$, then we have $x^7+x^2+x^{2^m+1}=0$ which implies that $x^{2^m}=x^6+x$. This leads to $x=(x^6+x)^{2^m}=(x^{2^m})^6+x^{2^m}=(x^6+x)^6+x^6+x$, i.e., $x^{20}+x^{10}+1=0$, a contradiction since $x^2+x+1=0$ has no solution in $U$ if $m$ is even. Then, for any $x\in U$, the polynomial $\phi_3(x^3)$ can be written as
\begin{eqnarray*}
  \phi_3(x^3)=x^3(1+x^{-2}+x^5)^{2^m-1}=\frac{x^7+x^5+1}{x^7+x^2+1}.
\end{eqnarray*}
Notice that $\gcd(x^7+x^5+1,x^7+x^2+1)=\gcd(x^5+x^2,x^7+x^2+1)=\gcd(x^3+1,x^7+x^2+1)=\gcd(x^3+1,x^2+x+1)=x^2+x+1$. Then, by a simple calculation, $\phi_3(x^3)$ can be reduced to
\begin{eqnarray*}
  \phi_3(x^3)=\frac{x^7+x^5+1}{x^7+x^2+1}=\frac{x^5+x^4+x^3+x+1}{x^5+x^4+x^2+x+1}.
\end{eqnarray*}

Assume that $\phi_3(x^3)=\phi_3(y^3)$ for some $x,y\in U$, then by a detailed but simple calculation one gets
\begin{eqnarray*}
  (x+y)[x^4y^3+x^4y^2+x^3y^4+x^3y^2+x^2y^4+x^2y^3+x^2y^2+x^2y+x^2+xy^2+x+y^2+y]=0.
\end{eqnarray*}
We next show that $x^4y^3+x^4y^2+x^3y^4+x^3y^2+x^2y^4+x^2y^3+x^2y^2+x^2y+x^2+xy^2+x+y^2+y=0$ has no solution $x,y\in U$. Let $y=ux$ for some $u\in U$, then it becomes
 \begin{eqnarray} \label{eq-u-4th}
 c_4u^4+c_3u^3+c_2u^2+c_1u+c_0=0,
 \end{eqnarray}
where the coefficients are given by
\begin{eqnarray*}
  c_4=(x^7+x^6),  c_3=(x^7+x^5),  c_2=(x^6+x^5+x^4+x^3+x^2),  c_1=(x^3+x), c_0=(x^2+x).
\end{eqnarray*}
Therefore, we need to show  \eqref{eq-u-4th} has no solution  $u\in U$ for any $x\in U$. Note that \eqref{eq-u-4th} has no solution in $U$ if $x=1$ and $u=1$ is not a solution to \eqref{eq-u-4th} for any $x\in U$.
Similar as the proof of Theorem \ref{thm-3q}, according to Lemma \ref{lem-U}, replace $u$ by $\frac{z+\gamma}{z+\overline{\gamma}}$ for some fixed $\gamma\in\ftwon\backslash\ftwom$, we then need to prove
\begin{eqnarray*}
 c_4(\frac{z+\gamma}{z+\overline{\gamma}})^4+c_3(\frac{z+\gamma}{z+\overline{\gamma}})^3+c_2(\frac{z+\gamma}{z+\overline{\gamma}})^2+c_1(\frac{z+\gamma}{z+\overline{\gamma}})+c_0=0
\end{eqnarray*}
has no solution $z\in\ftwom$. Since $x\in U$ and $x\ne 1$, taking $\gamma=1/x\in\ftwon\backslash\ftwom$ and multiplying $(z+\overline{\gamma})^4$ on both sides of the above equation, it gives
\begin{eqnarray*}
z^4+\frac{x^8+1}{x^4}z^3+\frac{x^8+x^6+x^2+1}{x^4}z^2+1=0.
\end{eqnarray*}
Replacing $z$ by $1/z$ and then multiplying by $z^4$, the equation becomes
\begin{eqnarray}\label{eq-4th-noroot}
z^4+\frac{x^8+x^6+x^2+1}{x^4}z^2+\frac{x^8+1}{x^4}z+1=0,
\end{eqnarray}
which has no solution in $z\in\ftwom$ for any $1\ne x\in U$ as shown in Lemma \ref{lem-3-25}. Then \eqref{eq-u-4th} has no solution  $u\in U$ for any $x\in U$.
This completes the proof. \hfill $\Box$

\begin{Lemma}\label{lem-3-25}
 Let the notation be defined as above. Then \eqref{eq-4th-noroot} has no solution in $z\in\ftwom$ for any $1\ne x\in U$.
\end{Lemma}
\Proof By Lemma \ref{lem-Leonard-W}, we first determine the number of solutions to $g(y)=y^3+a_2y+a_1=0$ in $\ftwom$, where $a_2=\frac{x^8+x^6+x^2+1}{x^4}$ and $a_1=\frac{x^8+1}{x^4}$.
For any $1\ne x\in U$, let $c=x+\frac{1}{x}\in\ftwom$, then $x, \frac{1}{x}\in\ftwon\backslash\ftwom$ are the two solutions of $z^2+cz+1=0$ which means that $\Tr_1^m(\frac{1}{c})=1$ according to Lemma \ref{lem-Quadratic}. Moreover, one can have
$$a_2=\frac{x^8+x^6+x^2+1}{x^4}=\frac{(x^8+1)+(x^6+x^2)}{x^4}=c^4+c^2,\;a_1=\frac{x^8+1}{x^4}=c^4.$$
Then, it can be readily verified that $g(y)=0$ can be rewritten as
\begin{eqnarray*}
(\frac{y}{c^2+c})^3+\frac{y}{c^2+c}+(\frac{c}{c+1})^3+\frac{c}{c+1}=0,
\end{eqnarray*}
which implies that $\frac{y}{c^2+c}=\frac{c}{c+1}$, i.e., $y=c^2\in\ftwom$ is a solution to $g(y)=0$. By a simple calculation, the polynomial $g(y)$ can be factorized as
\begin{eqnarray*}
g(y)=(y+c^2)(y^2+c^2y+c^2).
\end{eqnarray*}
Note that $\Tr_1^m(\frac{c^2}{c^4})=\Tr_1^m(\frac{1}{c})=1$ which implies that $y^2+c^2y+c^2=0$ has no solution in $\ftwom$ due to Lemma \ref{lem-Quadratic}. Therefore, $g(y)=0$ has exactly one solution $y=r_1=c^2\in\ftwom$, and then the desired result follows from Lemma \ref{lem-Leonard-W} and the fact that $\Tr_1^m(w_1)=\Tr_1^m(r_1^2/a_1^2)=\Tr_1^m(\frac{c^4}{c^8})=\Tr_1^m(\frac{1}{c})=1$. This completes the proof. \hfill $\Box$

\begin{Theorem}\label{thm-514}
 Let $n=2m$ satisfy $\gcd(5,2^m+1)=1$. Then the trinomial $f(x)$ defined by \eqref{def-PP} is a permutation if $(s,t)=(\frac{1}{5},\frac{4}{5})$.
\end{Theorem}

\Proof By $\gcd(5,2^m+1)=1$, similar as the proof of Theorem \ref{thm-3-14}, to complete this proof, it is sufficient to prove that $\phi_4(x^5)=x^5(1+x^{5s}+x^{5t})^{2^m-1}=x^5(1+x+x^4)^{2^m-1}$ permutes $U$, which is equivalent to showing
\begin{eqnarray*}
  \phi_4(x^5)=x^5(1+x+x^4)^{2^m-1}=\frac{x(x^4+x^3+1)}{x^4+x+1}
\end{eqnarray*}
permutes $U$ since $x^4+x+1\ne 0$ for any $x\in U$, as shown in the proof of Theorem \ref{thm-3q}.

Assume that $\phi_4(x^5)=\phi_4(y^5)$ for some $x,y\in U$, then one gets
\begin{eqnarray*}
  (x+y)[x^4y^4+x^4y+x^4+x^3y^2+x^3y+x^3+x^2y^3+x^2y^2+x^2y+xy^4+xy^3+xy^2+y^4+y^3+1]=0.
\end{eqnarray*}
We next need to prove that $x^4y^4+x^4y+x^4+x^3y^2+x^3y+x^3+x^2y^3+x^2y^2+x^2y+xy^4+xy^3+xy^2+y^4+y^3+1=0$ has no solution $x,y\in U$ with $x\ne y$. Let $y=ux$ for some $u\in U$ and $u\ne 1$, then it can be rewritten as
 \begin{eqnarray} \label{eq-u-514}
 c_4u^4+c_3u^3+c_2u^2+c_1u+c_0=0,
 \end{eqnarray}
where the coefficients are given by
\begin{eqnarray*}
  c_4=(x^8+x^5+x^4),  c_3=c_2=c_1=(x^5+x^4+x^3), c_0=(x^4+x^3+1).
\end{eqnarray*}
According to Lemma \ref{lem-U}, substitute $u$ by $\frac{z+\gamma}{z+\overline{\gamma}}$ for some fixed $\gamma\in\ftwon\backslash\ftwom$, we then need to prove
\begin{eqnarray*}
 c_4(\frac{z+\gamma}{z+\overline{\gamma}})^4+c_3(\frac{z+\gamma}{z+\overline{\gamma}})^3+c_2(\frac{z+\gamma}{z+\overline{\gamma}})^2+c_1(\frac{z+\gamma}{z+\overline{\gamma}})+c_0=0
\end{eqnarray*}
has no solution $z\in\ftwom$. If $x=1$, then \eqref{eq-u-514} is reduced to $u^4+u^3+u^2+u+1=(u^5+1)/(u+1)=0$ which has no solution in $U$ since $\gcd(5,2^m+1)=1$. For the case of $x\ne 1$, taking $\gamma=1/x\in\ftwon\backslash\ftwom$ and multiplying $(z+\overline{\gamma})^4$ on both sides of the above equation, it then gives
\begin{eqnarray*}
(x^2+x+1)^4 z^4+(x+1)^6(x^2+x+1)z+(x^4+x^3+x^2+x+1)^2=0.
\end{eqnarray*}
Notice that \eqref{eq-u-514} is reduced to $x(u^4+1)=0$ if $x^2+x+1=0$, which leads to $u=1$, i.e., we get the desired result $x=y$. If $x^2+x+1\not=0$, then the above equation can be written as
\begin{eqnarray}\label{eq-514}
z^4+(\frac{x^2+1}{x^2+x+1})^3 z+\frac{x^8+x^6+x^4+x^2+1}{x^8+x^4+1}=0.
\end{eqnarray}
Therefore, the desired result follows from Lemma \ref{lem-514} presented below. This completes the proof. \hfill $\Box$

\begin{Lemma}\label{lem-514}
 Let the notation be defined as above. Then \eqref{eq-514} has no solution in $z\in\ftwom$ for any $1\ne x\in U$ if $\gcd(5,2^m+1)=1$.
\end{Lemma}
\Proof In order to apply Lemma \ref{lem-Leonard-W}, we first consider the solutions to $g(y)=y^3+a_2y+a_1=0$ in $\ftwom$, where $a_2=0$, $a_1=(\frac{x^2+1}{x^2+x+1})^3$.
For any $1\ne x\in U$, let $c=x+\frac{1}{x}\in\ftwom$, then $x, \frac{1}{x}\in\ftwon\backslash\ftwom$ are the two solutions of $z^2+cz+1=0$ which means that $\Tr_1^m(\frac{1}{c})=1$ according to Lemma \ref{lem-Quadratic}. Further, we have
$$a_2=0,\;a_1=(\frac{c}{c+1})^3, a_0=\frac{x^8+x^6+x^4+x^2+1}{x^8+x^4+1}=(\frac{c^2+c+1}{c^2+1})^2.$$
Note that $\gcd(5,2^m+1)=1$ if and only if $m\not\equiv 2\pmod{4}$. We then can discuss it as follows:
\begin{enumerate}
  \item [1)] $m$ is odd: In this case, we have $\gcd(3,2^m-1)=1$ which means that $g(y)=y^3+a_2y+a_1=0$ has exactly one solution $y=r_1=\frac{c}{c+1}\in\ftwom$. Moreover, we have $\Tr_1^m(w_1)=\Tr_1^m(a_0r_1^2/a_1^2)=\Tr_1^m(\frac{c^4+c^2+1}{c^4})=\Tr_1^m(1+\frac{1}{c^2}+\frac{1}{c^4})=\Tr_1^m(1)=1$.
  \item [2)] $m\equiv 0\pmod{4}$: For this case $g(y)=y^3+a_2y+a_1=0$ has exactly three solutions $y_i=r_i=\theta^{i-1}\frac{c}{c+1}\in\ftwom$ for $i=1,2,3$, where $1\ne\theta\in\ftwom$ and $\theta^3=1$. Then, we have $$\Tr_1^m(w_i)=\Tr_1^m(a_0r_i^2/a_1^2)=\Tr_1^m(\theta^{2(i-1)}\cdot(1+\frac{1}{c^2}+\frac{1}{c^4})).$$
      Clearly, $\Tr_1^m(w_1)=\Tr_1^m(1+\frac{1}{c^2}+\frac{1}{c^4})=\Tr_1^m(1)=0$ due to $m$ is even. By the fact $\theta^3=1$ with $\theta\ne 1$, we have $\theta\in \mathbb{F}_{2^2}$ which implies that $\theta^4=\theta$, $\theta^2+\theta+1=0$ and $\Tr_1^m(\theta)=\Tr_1^{m/2}(\Tr_{m/2}^{m}(\theta))=\Tr_1^{m/2}(\theta\Tr_{m/2}^{m}(1))=0$. Then, $\Tr_1^m(w_2)=\Tr_1^m(\theta^2+\frac{\theta^2}{c^2}+\frac{\theta^2}{c^4})=\Tr_1^m(\frac{\theta}{c}+\frac{\theta}{c^2})=\Tr_1^m(\frac{\theta}{c}+\frac{\theta^4}{c^2})=\Tr_1^m(\frac{\theta}{c}+\frac{\theta^2}{c})
      =\Tr_1^m(\frac{\theta+\theta^2}{c}) =\Tr_1^m(\frac{1}{c})=1$. Similarly, we can obtain $\Tr_1^m(w_3)=1$.
\end{enumerate}
 Therefore, according to Lemma \ref{lem-Leonard-W}, we can conclude that \eqref{eq-514} has no solution in $z\in\ftwom$ no matter $m$ is odd or even. This completes the proof. \hfill $\Box$

In what follows, we list all the known pairs $(s,t)$ such that the polynomials of the form \eqref{def-PP} are permutations in Table \ref{Table1}. Note that the ``Equivalent Pairs" column in Table \ref{Table1} indicates that the pairs in this column also lead to permutation polynomials of the form \eqref{def-PP} only if the corresponding values modulo $(2^m+1)$ exist and the conditions in the second column are satisfied. Take $(s,t)=(3,-1)$ and its an equivalent pair $(\frac{3}{5},\frac{4}{5})$ for an example, it is shown in Theorem \ref{thm-3q} that $f(x)$ defined by \eqref{def-PP} with $(s,t)=(3,-1)$ is a permutation polynomial if $m$ is even. Then, by Lemma \ref{lem-equiv}, we can claim that $f(x)$ defined by \eqref{def-PP} with $(s,t)=(\frac{3}{5},\frac{4}{5})$  is also a permutation polynomial if $m$ is even and $\gcd(5,2^m+1)=1$.
\begin{table}[ht]
\caption{Known pairs $(s,t)$ such that $f(x)$ defined by \eqref{def-PP} are permutation polynomials}\label{Table1}
\begin{center}{
\begin{tabular}{|c|c|c|c|}
  \hline
  $(s, t)$ & Conditions & Equivalent Pairs & Reference \\ \hline\hline
  $(k, -k)$  & see Theorem \ref{thm-known pairs}   &  $(\frac{k}{2k-1},\frac{2k}{2k-1})$, $(\frac{k}{2k+1},\frac{2k}{2k+1})$ & \cite{Ding-QWYY} \\ \hline
  $(2, -1)$ &  positive $m$  & $(1,\frac{1}{3})$, $(1,\frac{2}{3})$ & \cite{Ding-QWYY}  \\ \hline
 $(1, -\frac{1}{2})$ &  $m\not\equiv 0\pmod{3}$  & $(1,\frac{3}{2})$, $(\frac{1}{4},\frac{3}{4})$ & \cite{Li-Qu-Chen}  \\ \hline
  $(-\frac{1}{3}, \frac{4}{3})$ &  $m$ even & $(1,\frac{1}{5})$, $(1,\frac{4}{5})$ & Theorem \ref{thm-3-14}   \\ \hline
   $(3, -1)$ &  $m$ even & $(\frac{3}{5},\frac{4}{5})$, $(\frac{1}{3},\frac{4}{3})$ & Theorem \ref{thm-3q}   \\ \hline
    $(-\frac{2}{3}, \frac{5}{3})$ &  $m$ even & $(1,\frac{2}{7})$, $(1,\frac{5}{7})$ & Theorem \ref{thm-3-25}   \\ \hline
     $(\frac{1}{5}, \frac{4}{5})$ &  $\gcd(5,2^m+1)=1$ & $(1,-\frac{1}{3})$, $(1,\frac{4}{3})$ & Theorem \ref{thm-514}   \\ \hline
\end{tabular}}
\end{center}
\end{table}

Notice that any pair  $(s, t)$ such that $f(x)$ defined by \eqref{def-PP} is a permutation polynomial can produce many permutation trinomials by using Lemma \ref{lem-Zieve}.  According to Theorem \ref{thm-3-14}, Theorem \ref{thm-3q}, Theorem \ref{thm-3-25} and Theorem \ref{thm-514}, the following result can be readily verified based on Lemma \ref{lem-Zieve}.

\begin{Corollary}
Let $n=2m$, $q=2^m$, and $k$ be positive integers with $\gcd(2k+1,q-1)=1$. Then the following trinomials are permutations over $\ftwon$:
\begin{enumerate}
  \item [1)] $x^{(q+1)k+1}+x^{(q+1)k+(2q^2-q+2)/3}+x^{(q+1)k+(q^2+4q-2)/3}$  if $m$ is even;
  \item [2)] $x^{(q+1)k+1}+x^{(q+1)k+3q-2}+x^{(q+1)k-q+2}$  if $m$ is even;
  \item [3)] $x^{(q+1)k+1}+x^{(q+1)k+1+(q-1)^2/3}+x^{(q+1)k+(2q^2+5q-4)/3}$  if $m$ is even;
  \item [4)] $x^{(q+1)k+1}+x^{(q+1)k+1+l(q-1)}+x^{(q+1)k+1+4l(q-1)}$, where $5l\equiv 1\pmod{q+1}$.
\end{enumerate}
\end{Corollary}


To end this section, we propose two problems about the permutation trinomials with the form \eqref{def-PP}, which will cover most of the known cases if they are solved.

\begin{Open}
Determine the condition on $s$ such that $f(x)$ defined by \eqref{def-PP} is a permutation polynomial for $s+t=1$.
\end{Open}

\begin{Open}
Determine the condition on the integer $k$ such that $f(x)$ defined by \eqref{def-PP} is a permutation polynomial for $(s,t)=(2k,-k)$.
\end{Open}

\section{Concluding Remarks}

In this paper several new classes of permutation trinomials over $\ftwon$ were obtained from Niho exponents based on some techniques in solving equations with low degrees over finite fields. It should be noted that the presented permutation trinomials possess the form \eqref{def-PP} with Niho exponents and thus it can be readily verified that they are multiplicative inequivalent to the known ones listed in Theorem \ref{thm-known}. Computer experiments show that there still have many permutation trinomials of the form \eqref{def-PP}, it will be nicer if more such explicit permutations can be obtained in a general way.

\section*{Acknowledgements}

The authors would like to thank the editor and reviewers for their comments that improved the quality of this paper. This work was supported by the Norwegian Research Council.

\end{document}